\documentclass[reqno]{amsart}
\usepackage{graphicx}
\usepackage{hyperref}

\voffset = - 0.5 cm
\hoffset = - 2 cm
\begin{document}

\title{Footprinting in a course on energy}

\author{Seth A. Major
}

\date{24 July 2023}
\address{Department of Physics\\ Hamilton College\\ Clinton NY 13323 USA}

\email{smajor@hamilton.edu}

\begin{abstract}
Footprints provide a way to estimate the relative impact of processes and products on the global climate. Including footprint analysis in a course on energy simultaneously provides students with an understanding of this tool and a quantitative guide to approaches that address climate change. College-level classroom activities for (primarily) process-based life cycle carbon footprinting are discussed. 

\end{abstract}

\maketitle


\section{Introduction}

Whether teaching upper-level statistical mechanics or an introductory course on energy, I find that students are keen to learn ways we can address climate change.  Footprinting is one quantitative method to guide approaches to reducing emission of anthropogenic greenhouse gases.  In this article I share a `nugget', as described in the prompt for the special collections on the physics of the environment, sustainability, and climate change: a brief description of classroom activities on footprinting to which the students responded favorably.   

In the next section I sketch the course settings in which I have implemented footprinting activities, including discussion of the necessary background required for the activities.  While the discussion focuses on footprinting in a course on energy, these activities may be included in courses statistical physics courses. In section \ref{footprint} I briefly introduce carbon and ecological footprints. The classroom footprinting activities are described in section \ref{classroom}.  Finally, I conclude with some discussion and student feedback to the activities in section \ref{discussion}.

\section{Teaching contexts}

Elements of footprinting can be most easily incorporated into college physics courses on energy, but they can also be incorporated into traditional upper-level statistical physics courses.  The classroom activities described in the next section were developed for ``Humanity's Global Impact and the Adirondacks" (Coleg-370). This course was part of the Hamilton Adirondack Program in 2016 and included an approximately two-week development of footprint analysis.  See Appendix \ref{descrip} to more on the program and course.  Sample example questions are given in Appendix \ref{questions}.

The goals of Coleg-370 are:
\begin{enumerate}
\item To understand a detailed ``back of the envelope" calculation of global warming due to CO$_2$ emissions from fossil fuels.  
\item To discover insights of the system-wide thinking of thermodynamics. 
\item To become more number savvy. 
\item To be able to determine the reduction of CO$_2$ emissions due to specific actions or policies.   
\item To practice using those tools by applying them to local contexts.
\end{enumerate}
We use footprinting to understand the flow of carbon in producing items or running activities. Applying the quantitative and spreadsheet skills developed earlier in the course, students address what factors might be important in carbon production and find gaps in our understanding and data. They see how footprints can be used to help direct changes to address climate change. 

The course covers necessary background for the footprinting activities including, (i) the physics of energy, taught in an essentially traditional approach, (ii) quantitative skills in building spreadsheets and in incorporating uncertainties taught in laboratory exercises, and (iii) a quantitative ``back of the envelope" calculation of climate change that includes the carbon production in burning fossil fuels and an estimation of the global warming due to this additional CO$_2$. The approach to the calculation is similar to Tom Murphy's ``Do the Math" blog \cite{tom_blog}).\footnote{I also include this curricular element in my junior-level statistical physics courses as an application of blackbody radiation.} Before concluding the semester with the footprinting, we discuss energy systems.\footnote{Valuable for this portion resources include MacKay \cite{mackay} (excellent and free, but focused on the UK) and Murphy \cite{tom_book}} In upper-level statistical mechanics courses students have much of this background already so I only include (iii) in this context.

To support the carbon accounting, it is helpful to find the carbon intensity of combustion and motivate the unit CO$_{2e}$. In the calculation of carbon produced in burning fossil fuels, one finds that the carbon intensity of fossil fuels is about 3g CO$_2$/g (grams of CO$_2$ emitted per gram of combusted fuel) across all fossil fuels.  In the course we investigate the relative radiative forcing of greenhouse gases by comparing absorption around 10 $\mu$m, which is the peak wavelength of a blackbody spectrum at about $T= 288$ K. This gives a metric of the relative radiative forcing among greenhouse gases. Because the radiative forcing and the lifetimes of these gases in the atmosphere vary significantly, it is helpful to normalize the impacts.  We, thus, come to the standard carbon footprint unit of grams of CO$_{2}$ equivalent or CO$_{2e}$: the amount of CO$_2$ that would have the equivalent global climate impact over a hundred year period.\footnote{The radiative forcing of key greenhouse gases is both highly variable and complex (our understanding is evolving, see \cite{methane_update}). For instance, by integrating absorption spectra in the infrared we see that methane is about 60 times more potent than CO$_2$. But methane chemically breaks down faster in the atmosphere. Over the hundred year period it is about 28 times more potent than CO$_2$.  For more see the IPCC report \cite{IPCC}, Chapter 9 in \cite{tom_book}, and \cite{MBL}.}


\section{Footprints}
\label{footprint}

Humanity's impact on the world is both broad and fine-grained. For example, it has a significant affect on biodiversity, availability of fresh water, soils, and climate, only some of which may be directly addressed by changes guided by footprinting.  When addressing climate change, it is natural to study the carbon footprints of products and processes to identify drivers in greenhouse gas emission.  Other accounting methods, such as the  ecological footprint (EF), widen the focus to include water and land areas to support activities. While both carbon and ecological footprinting techniques are still under development, the accounting nonetheless provides a useful quantitative guide to determine which avenues might reduce greenhouse gas emissions and, in the case of EF, our broader impacts on the environment. Other methods of analysis and modeling are necessary to guide choices in addressing other areas of our impact such as biodiversity loss. In this section carbon and ecological footprints are defined and sources are discussed. Activities based primarily on the carbon footprint metric are described in the next section.

Carbon footprints are usually based on a process-based life cycle assessment, an input-output analysis, or, a combination of both.\footnote{See the ``Where do the numbers come from?" chapter in \cite{MBL}.}  The process-based assessment accounts for the carbon and other greenhouse gases released in producing, using, and disposing of a product or completing an activity.  It is a bottom-up approach, building the total footprint from footprints of sub-processes. An advantage of this approach is that it is specific to the item. The carbon footprints of sub-processes are modeled so the origin of the carbon in the model is clear. But process-based assessment is subject to truncation error when sub-processes are inadvertently left out.  For example, a process-based carbon footprint of a smartphone could include emissions from manufacturing and transporting but might omit emissions due to mining the required raw materials.

The top-down approach of the input-output model\footnote{At its heart the work is nifty bit of linear algebra that tracks the outputs of one sector of industry with the inputs of other sectors.  The method was created by Wassily Leontief, for which he received the Nobel Prize in economics in 1973.} starts with the whole economy and its carbon output. As such, the input-output approach isn't subject to the truncation `leaks' in the process-based approach. However, input-output analysis does not have the specificity of process-based assessment. In part because the input-output is based on linear algebra and requires large data sets, I have not used this method in classes other than to refer its results during discussions of more complex processes such as the carbon footprint of iPhone production. As I describe in the next section, there are enough subtleties in process based assessment that it make sense to focus student work on the process-based approach in an introductory setting.

One engaging source on carbon footprinting is Berners-Lee's {\em The Carbon Footprint of Everything} \cite{MBL}.  Now in its second edition - it was previously titled {\em How Bad are Bananas?} - it makes for surprisingly fun reading. The notes in the back of the book are a goldmine for those wishing to look into footprinting methods and literature. One of the students in Coleg-370 wrote, ``The best way [to] overcome [the] barrier to action would be to have everyone read Berners-Lee." 

The ecological footprint (EF), developed by Wackernagel and Rees \cite{eco_foot}, recasts the environmental impacts of our activities (and populations) in terms of area of land and ocean required to generate the resources and to absorb the waste they generate.  The EF unit is numbers of earths, that is the number of earths to sustainably maintain a way of life.  This may be stated as ``If everyone on earth adopted the same lifestyle as those living in country $X$ then this would require $Y$ earths." (Typically, $Y>1$.) Students noted the appeal of this choice of expressing the result, ``... for me, footprints expressed in earths are the most meaningful, because it is a unit we can all visualize," as one student noted. 

The EF is extensively developed by the Global Footprint Network (GFN) \cite{GFN}. Useful sources include the online calculator \cite{eco_foot_online} and curricular materials available through EUSTEP (Enhancing Universities' Sustainability Teaching and Practice through Ecological Footprint), a partnership between the GFN and four universities \cite {eco_ed}.  

Although the EF has an intuitive unit, the computation of the EF is often country-based and, particularly with online calculators, can seem like a black box.  For instance, I assign students to use this calculator for their own footprints in two different countries. They often find that with the same personal inputs, the resulting ecological footprints are radically different, which initially puzzles students.  Even after they understand society's role in the variation of ecological footprint, it is not clear which policy choices lead to the significant differences.

\section{Footprinting in the classroom}
\label{classroom}

Students are ready to footprint when they have background in energy and the `back of the envelope' calculation of climate change and have some experience working with data, uncertainty, and spreadsheets. In this section I describe the classroom activity which includes warmup footprint calculations in the small-class setting of Coleg-370.  Afterwards I comment on changes that I have made to incorporate simple footprinting into a larger class setting in a statistical physics course. These activities are not prescriptive and may be adapted to different courses and settings. 

The footprint activity starts with guided discussion on an item's footprint that is (at least initially) uncomplicated. For instance, I hold up a cup of tea and ask, ``What is the carbon footprint of a mug of hot water?"\footnote{See \cite{DPB} for an approach using an electric kettle to explore energy use and life cycle assessment.} In a small class setting a whiteboard or a blackboard work well to keep track of the scope and to perform initial calculations.  Although I don't explicitly say this at the beginning, the initial round of discussion and calculation is really about the choices involved in defining the footprint.

I let students take a lead on the initial quantitative discussion, with my role being more of a scribe and an occasional guide.  My students tend to gravitate to a simple thermodynamical ``$Q=mc \Delta T$" computation of the energy required to boil the water.  For instance, we might calculate the carbon produced when burning natural gas to raise the temperature of tap water to boiling. We find a preliminary carbon footprint in terms of grams of CO$_2$ using the carbon intensity we found during the ``back of the envelope" climate change calculation.

Even during the initial discussion, students often find the complexities involved in even the most simple computations, but, if they do not, I push on the assumptions. Suppose we find an unrealistic footprint based on heating on a gas burner (about 4 g CO$_2$). I point out the assumption - often not explicitly stated -  that {\em all} the heat of combustion was transferred to the water. I prompt students to re-consider,  asking, ``If this was the actual footprint wouldn't we be comfortable holding our hand above the pan while the water is heating up?"  
This observation returns us to another round of quantitative discussion, defining the problem with more care, and refining the computation. It is also useful at this stage to look at more than one method of heating. Depending on whether there is equipment available - in Coleg-370 the kitchen was in an adjacent room - the next stage of taking data can be completed during the first class activity on footprinting. 

Equipped with timers, information on the stoves, and electrical consumption meters like `Kill-A-Watt' meters they boil water to refine their initial calculations. They time how long it takes to actually boil a cup of water on a burner of known power. During these measurements I encourage students to observe and record what they actually do. For example when boiling water on a stove one frequently uses a larger volume than one cup while when heating water for tea in a microwave one often heats just the water we intend to drink.\footnote{Thanks to Beth Parks for this observation.} For heating with an electric kettle, we use the local electrical mix on the grid from energy generation portion of the course.

At the end of these preliminary footprints, we return to the definition of the footprints and the manner in which we communicate the results with their limitations.  Now with students more aware of the processes that may be included, we discuss whether to include the footprint of potable water and waste systems. This reflection may lead us to new definitions, calculations, and results. For example, we might choose to find the carbon footprint of the energy required to heat water with natural gas, about 14 g CO$_2$, and to neglect the footprint of the infrastructure required to deliver the potable water. Likewise heating a cup of water using an electrical kettle, powered by the local grid, has a footprint of about 19 g CO$_{2e}$.\footnote{The change in units is due to the use of the electrical mix from the grid but the results may be directly compared due to the choice to normalize the impact using CO$_2$.}

The process of reflecting on the definition helps students see that variability in the scope of footprints. In most of these questions the goal is not only the ``correct answer", but rather the consciousness of the decisions that define our results. This is a key element of the approach to footprinting advocated here. Like some undergraduate physics laboratories the activity is a cyclical process of defining, calculating and measuring, reflecting on the results. The activity highlights the choices and limits involved in process-based carbon footprints.  In addition, by explicitly working through computations students have full access to what is done in these computations, unlike the results of an online footprint calculator.

It is possible to adapt this activity to other class settings, as I have done in a statistical physics course. In larger class settings, the initial activity can be done with group work using large, shared whiteboards for tables. (The physics department has cut $40 \times 90$ cm$^2$ whiteboards for such discussions.) Depending on the size of the class, this may require TA's to help guide the discussion. In my classes at least some students have access to kitchens. This measurement portion is completed as homework in `dorm-kitchen labs' and discussed in the next class meeting.  As I have adapted the activity in a more conventional instructional setting, the activity is a discussion taking less than one class period followed by homework activities and an in-class discussion during which the final versions are found. In statistical physics courses, which lack the coverage of energy systems included in Coleg-370, the renewable mix on the local grid and how much energy is lost in delivery are briefly discussed in class.

The first full footprint of a product starts with an item that (at least some) students know well so they can apply their understanding of the processes involved.  A main goal for this activity is a more complex footprint calculation. We explore the scope of what one might consider. We take more care to note where numerical inputs are `squishy', i.e. have high relative uncertainty, when different sources offer significantly different results, and/or when the data is from a simple internet search.  In addition to the blackboard it is very helpful to have cloud-shared documents and spreadsheets such as those hosted on DropBox, Google Drive, or OneDrive. This way students can quickly share tips, work, and sources during the discussion.  Additionally, it is helpful to have access to a high quality published study with plenty of direct measurements, allowing one to dig into details (such as \cite{lettuce} used in the example below).

For example in the HAP everyone made food and participated in the Essex Farm CSA food pickup. Some students interned on farms cultivating vegetables and were familiar with plastic row covers, harvesting, refrigeration, tractor use (and parts required for repairs), and irrigation.  So a natural choice was a farm product they knew well.  I chose a computation of the footprint of lettuce.  This first footprinting activity started with a discussion of contributing processes, dividing them into categories of ``to the farm gate", ``to the store door", and `to the dinner table". While we eventually studied all of these categories, we started with the ``to the farm gate" footprint.  A spreadsheet with the computation is available as supplementary material \cite{sup}. 

To manage the problem, and to clearly mark what is included in the footprint, it is essential to break down the product or process. In the initial discussion we used a blackboard to record the breakdown of individual processes involved asking, ``How is the product made? Where to the parts or inputs come from? What do we include in the footprint?" The resulting breakdown allows us to list smaller, manageable steps, to note needed data, and to define the calculation. The process is highly customized to the product or process.  In fact this is a strength of the process based assessment.

In the case of lettuce we asked, ``Was the lettuce grown outside or in a green house? Was the crop conventionally grown or organically grown? What is the tractor use? Is fertilizer used? Was there post harvest cooling? What sort of packaging, if any, was used? How was the produce transported? Do we need to include warehouse infrastructure?" The guides for the discussion were both the students' experience and the published study \cite{lettuce}. We discussed ways to express the result. For instance, what is the best unit to describe the footprint, an acre, pound, calorie, head, or serving of lettuce? The students chose a head since it represents both a common unit for the farm and the consumer.

Since 80\% of lettuce consumed is conventionally grown, in class we decided to find the carbon footprint of a conventionally grown head of lettuce as it leaves the farm. (Later, we compared the result to the footprint of lettuce produced locally.) Defining this footprint and determining which data we needed was conducted in a class discussion using both the blackboard and the shared spreadsheet.  During discussion students started to fill in the footprints for individual processes such as tilling, working from knowledge of what is required, adapting the results in the published study \cite{lettuce}, and checking basic inputs such as the rate of diesel consumption.

Next we extended the result to the carbon footprint of a conventionally grown head of lettuce as arrives prepared on the dinner table. Developing these footprints allowed us to find significant contributions to the carbon emissions of growing and delivering lettuce.  Key possible contributions included air freight, refrigerated transport, and greenhouse heating. The computations were done in teams focused on different individual processes using the cloud-shared spreadsheet. Work was done both in class and in homework. We included comments and citations of the sources in the spreadsheet so we could both check our work and use the spreadsheet as a reference for future footprints.

While the focus of Coleg-370 was not on society and policy, HAP did have such a focus. Reflecting on these results at the end of the fall harvest led to far ranging conversations. Discussions included our choice of diet and an apparent rural-urban split in the expectation of readily available lettuce year round versus other produce such as cabbage or pickled vegetables. Given the fossil-fuel-light farming methods, including horses (``shaddowfax" in MacKay's parlance) used at Essex Farm we realized that the literature did not contain data on the carbon production and sequestration on diversified farms such as Essex Farm, which suggested new research collecting the data. 

Other early footprinting activities might include on-campus initiatives such as end-of-year furniture swaps (comparing furnishing rooms with new sofas, etc. versus used furnishings from the swap), student-run coops, real-time energy monitoring systems, and on-campus composting. Ideally, these are student-driven or student-inclusive initiatives so that the students have knowledge and ownership of the systems and processes. Once students have found several footprints, and have access to a small library of examples, working independently becomes possible. Students then have some basis to evaluate whether their footprints make sense. 

In addition to footprints described in assigned reading in Berners-Lee \cite{MBL} we also studied carbon footprint analyses of iPhones (leaning heavily on Apple's version, e.g. \cite{foot_iphone}, and Berners-Lee's critique \cite{MBL}), solar panels, the 2016 Olympic Games in Rio \cite{rio}.  We created footprints for other locally produced goods, Hamilton College, and the Hamilton Adirondack Program (see Appendix \ref{adk_foot}) in more detail. For the HAP assessment we worked as a team, dividing the tasks of assessing housing, food, the guest speaker program, vehicle use, and outdoor activity gear. By this point in the semester it was not surprising to see that travel was the main contributor of carbon in the HAP footprint (Appendix \ref{adk_foot}). This led to suggestions on how to change the program structure to lower the carbon footprint. At the end of the footprint portion of the course, students were able to independently complete process-based carbon footprint calculations. Every student completed a new footprint as part of the final exam. 


Faculty preparation for the footprint activity includes running a few of these footprint analyses in advance. It is useful to work from published work with plenty of direct measurements and to keep open to new possibilities that students raise.  One of the pleasures of teaching this subject is learning from student insights. 

\section{Discussion}
\label{discussion}

Saadia Zahidi of the World Economic Forum is quoted as saying, ``What you can measure, you can address" \cite{Zahidi}. Footprinting offers a metric to prioritize solutions that address climate change. The activity also serves as a bridge between online `your footprint' calculators and summaries such as in \cite{MBL}, and full-blown, careful studies, making the process more transparent and instilling an understanding of its power and limitations.  
The footprinting activity requires background in energy and the basis of climate change, as well as a facility with spreadsheets. For a first footprint, it was helpful to choose an item that the students knew well, that wasn't too complicated, and that had at least one solid study to draw from. In Coleg-370 over the course of two weeks students developed a handful of process-based life cycle assessment carbon footprints of products and activities. In upper-level statistical mechanics courses, simple carbon footprints can be included as in-class example and in homework, although this omits the significant lessons on scope and definition that a longer unit includes.

Students in both these settings responded enthusiastically to the inclusion of discussion of climate change and, in the case of Coleg-370, to the ``What can we do about it?" unit based on footprints.  Perhaps part of the reason is due to the surprises: 
Purchased out of season air-freighted fruit? You have bought a ``big foot", a footprint many times larger than similar, locally produced food.\footnote{It is rarely simple to determine whether a product traveled by air.  However, stores often include country of origin. Perishable foods that travel far are often air-freighted. For example, fresh blueberries from Peru purchased in the Adirondacks in December will carry a large foorprint due to air travel.} Considering a change in diet? A vegan diet can have as big a carbon footprint as a meat-based diet, if the food travels by air. Computing the impact through footprints offers a guide to change, whether it be national transportation policy or sealing leaks in houses. The footprinting activity offered students a quantitative guide to changes to address climate change. Some lifestyle changes such as night setback temperatures or different methods to cook pasta were easy to implement. We can draw from more comprehensive plans for greenhouse gas reduction (see, for instance, \cite{drawdown}) for potential larger scale changes. 

In Coleg-370 assessment was through written feedback, and focused on the course goals (4) and (5). The students welcomed the footprinting work as a guide to what might be addressed to reduce carbon emission. For example, in reflecting on the footprinting activity one student wrote,
``I know that learning how much a footprint flying, driving, and heating have has totally changed my perspective.  I think having footprint numbers on things like boarding passes, gas pumps, and heating bills could really help to make people rethink their usage. ... this information needs to be part of every day to day activity..." 
Implementing footprinting activities in physics courses creates opportunities for such a change in perspective.  
\bigskip 

\noindent{\large\bf Acknowledgements} Many thanks to the students in the Hamilton Adirondack Program, Janelle Schwartz, and Hamilton College for support. Thanks also to the anonymous reviewers whose extensive comments helped to shape the article.

\appendix

\section{Coleg-370 Course detail}
\label{descrip}

In 2016 I was fortunate to teach in the Hamilton Adirondack Program (HAP). This innovative program founded by Janelle Schwartz 
blended coursework in the Adirondack park with a deep engagement with the environment and the local community.  The HAP offered students a wide breadth of experiences from academics focused on both global concerns and the local community to food production and preparation to working in the area through internships.

The idea was to place global thinking on a quantitative foundation and then apply this thinking to solutions on a local level. To create the syllabus for the course I worked backwards from a `back of the envelope' calculation of climate change and footprinting. This climate change calculation is similar to what is done in Chapter 9 of \cite{tom_book} and the blog \cite{tom_blog}, both by Murphy.  In preparation of this article I updated the teaching resources. Asking, ``What background is required to understand the calculation?" and  ``What skills are required for the footprinting?" pointed toward material content including the physics of energy and light and basic chemistry. But the questions also pointed to necessary background in spreadsheets, heinous unit conversions, and simple quantitative analysis. I included earlier activities to introduce and practice these skills. Laboratories were tailored to the site and integrated into the weekly assignments. In addition to energy labs we carefully modeled the heat flow in residential structures including blower door testing on the campus structures. In another (and favorite) lab, students built their own `homes', small structures of their own design and construction. The students tested the heat loss of their structures using electrical resistive heat and temperature loggers.

In the first part of the course I followed the `force-work-energy' path towards understanding energy, essentially following standard physics texts. Although I followed a philosophy of `choose your favorite source', I assigned reading by topic in Bloomfield's ``How Things Work" \cite{bloomfield} and the algebra-based Open Stax College Physics \cite{stax}. Additional useful texts include Muller's ``...for Future Presidents" series e.g. \cite{muller}, Murphy's {\em Energy and Human Ambitions on a Finite Planet} \cite{tom_book}, and Wolfson's {\em Energy, Environment, and Climate} \cite{wolfson}. The physics background finished with the basics of thermodynamics with it's benefit of system-wide macroscopic thinking.

At this stage the students' work was a mix of qualitative, quantitative, and working-with-numbers questions (significant figures, basic uncertainty, agreement, unit conversion). They also built skills working with spreadsheets. Example questions are in Appendix \ref{questions}. The tailored approach laid the foundation for footprinting. We studied sources of power, locally, nationally and globally, electrical grids, electrical energy mix.  Although focused on the UK, MacKay's {\em Sustainable Energy without the Hot Air} \cite{mackay} is an excellent source for this material. Murphy's book offers a most recent presentation more focused on North America \cite{tom_book}. The course also included a field trip to visit to a local small-scale hydropower plant and guest speakers. 

Although some students in Coleg-370 had extensive undergraduate exposure to physics, most of the students said that they would not have taken a physics course on the Hamilton College main campus. It was rewarding to observe their success and valuable for all of us to hear different perspectives, some of which were quite different than I usually have in the physics classroom.  Grading was based on weekly homework questions, computations and short essays, into which lab writeup incorporated, a midterm, and a final. 


\section{Example questions}
\label{questions}

This is a sampling of questions\footnote{These questions are from 2016.} from the homework, quizzes, and labs.  
\begin{itemize}
	\item Do you do work while kneading bread? Explain your answer. [Students had just started to make bread for the community.]
	\item  Explain why we often build structures with steeper roofs in the mountains than in the lowlands far from snowfall. 
	\item Water descends in Gulf Brook from the Mountain House to the Ausable River in Keene.  
	What is the maximum electrical energy that the water could generate in a day?  Assume a flow of 911 kg per day. Hint: You need one more bit of information which is available on the topo map. Compare this to the electrical energy consumed in one day on site.
	\item By now you have done a lot of reading and calculating on hydro power.  What role can hydro power play in the Adirondack region electrical energy mix? Draw from our conversations with Matt Foley [owner of small scale hyrdo facility] in Westport, 
	the EPA feasibility study, the previous problem, and your reading. Make sure that, in MacKay's words, your solution ``adds up". 
	\item  Find the carbon and ecological footprints of (a) 1 pint of blueberries (375 g) flown to Newark, NY from Lima, Peru and trucked to the local Hannaford's. You may want to focus on the major contributing factors;  
	(b)  1 pint of raspberries grown at Essex Farm [the program's CSA]; and
	 (c) 1 pint of Essex Farm raspberries frozen for 3 months - this is to mimic the out of season nature of (a).  Comment on the results. For yields use MOFGA's estimates: ``... typical yields are 4,000 to 5,000 pounds [per acre]."
	  \item  Jenkins has a sustainable energy solution for the Adirondack Park \cite{jenkins}.  Summarize the solution and compare it to the scenarios in MacKay.  Why are they so different?  What are some advantages and disadvantages of Jenkins' solution?  (Does the park `belong' to people living in it? To NY state? To all? Are there benefits to such local energy solutions?)
	\item Congratulations!  You have been appointed to the Hamilton College's  `Blue Sky' Committee. 
	(This body actually existed, although the official title was the ``Imagining Hamilton Committee".)  
	As the climate expert on the committee, you are tasked with developing a plan to accelerate Hamilton College's Climate Plan so that the college is climate neutral by 2030.
	\begin{enumerate}
		\item[a.] Estimate the current carbon footprint of Hamilton College. For data, use Berners-Lee and Hamilton College's FY 2016 primary energy, $2.40 \times 10^{6}$ kWh of electricity and $1.14 \times 10^6$ therms of natural gas.  Do not include student travel to and from the campus. Develop a spreadsheet for Hamilton College's total carbon footprint. 
		\item[b.] Use MacKay, Jenkins, and Berners-Lee to map out a climate neutral solution. Include the possible benefit of the forest on the 1280 acre campus, about 800 of which is forested. At this stage do not worry about financial constraints. 
		\item[c.] Now discuss the feasibility of your plan. 
		\item[d.] Submit an electronic version of your spreadsheet.
	\end{enumerate} 
	\item What does Berners-Lee's input-output model give for a minimum emission for spending \$1?  Ponder this for awhile then give the argument for intentionally moderating one's consumption, i.e. spending those dollars.

	\item Re-run the numbers for a recently installed 3.92 kW system that cost \$4100 (installed, state and federal incentives included).  Please find the expected annual energy savings, payback period, the return on investment, `financial impact of solar panels', and the CO$_2$ payback period.  Assume a cost of electricity of \$0.14/kWh, and the upstate electrical mix of 51, 27, 16, 4, 2 percent for gas, nuclear, hydro, other renewables, and coal, respectively (www.eia.gov for New York in July). Comment on this system and the example Berners-Lee uses.
	
	\item The director of Physical Plant sends out an annual `Building Heating' email  - perhaps you have seen it - writing, ``make sure windows are closed in the cooler evening hours."  Still many windows are open around campus during cold weather.  To increase participation, what would you add to this email on how windows operate? Many of us have a tendency - I think - to view this email as a distraction.  How might the director circumvent this problem?
\end{itemize}

\section{Hamilton College's  Adirondack Program Footprint}
\label{adk_foot}

\begin{figure}[h] 
\begin{center}
\includegraphics[scale=.7,angle=0]{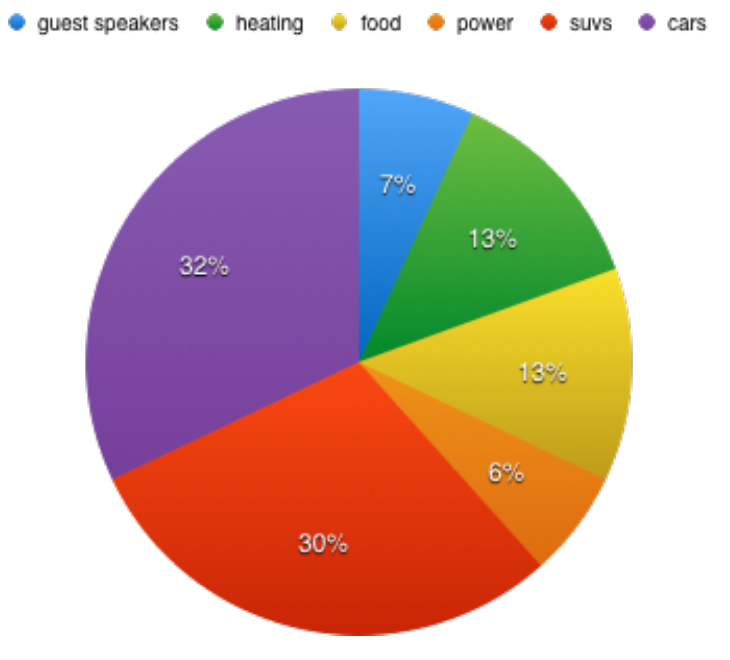}
\caption{\label{figure} The main contributions to the carbon footprint of the Fall 2016 HAP program.  The label `cars' refers to the use of student vehicles for all travel during the fall, but not to and from campus; `suvs' represents the footprint of the program vehicles based on actual mileage; `food' represents our estimate for the footprint of food purchases; `heating' is a rough estimated footprint for heating the --- site buildings based on New York averages; `guest speakers' refers to the footprint of the travel to and from the site, adjusted for multipurpose trips; and `power' is the carbon footprint of the electrical energy consumed based on energy mix in the [local area].}
\end{center}
\end{figure}

\noindent {\bf Program Footprint:}  One of the themes of the course was to answer ``What can we do about climate change?" As part of the focus on the local, we studied the Hamilton Adirondack Program's carbon footprint and estimated it to be about 4 tons of CO$_{2e}$ per full-time community member per semester, see Fig. \ref{figure}.  This was defined through carbon accounting for food, heat, electrical use, transport during the semester (but not to and from the campus) for the full community. It did not include contributions from infrastructure such as the campus buildings and methane emissions from composting, See below for more discussion.

It is easy to see that transportation dominates the program's footprint. The transportation (personal ``cars" and program ``suv's") results from the structure of the internships, food sourcing, and recreational activities. If reducing the carbon footprint of the program is a goal, reducing the transportation portion is probably the single most effective step. This can be obtained for instance by clustering program internships in the Keene - Lake Placid area so that students can both carpool more and can travel fewer miles.

The chief sources of uncertainty in the Program's footprint come from relatively poor heating and food production data. Although our landlord was generous with access to the energy suppliers, ultimately the --- heating data was not complete enough to use. We used average New York numbers for heating. In any case since the program extends only about one month into the heating season, this contribution to the footprint is not large.  The other big gap in our data is the footprint of Essex farm (and other local) food. This is highly uncertain and we used a mix of US and UK studies in our estimate. 

It is early days yet for the carbon footprinting of colleges and universities is in its infancy and comparisons may be fraught with complications.  The location of a campus plays a large role in the footprint due to heating and/or cooling. Nevertheless as a point of reference, one careful study of Lancaster University found about 2.5 tons of CO$_{2}$e per staff and student member per semester.  
(The Higher Education Funding Council for England has set carbon reduction targets.)   


\bigskip

\noindent {\bf Author Declarations}

The author has no conflicts to disclose.


\end{document}